\documentstyle[11pt,newpasp,twoside,epsf]{article} 

\begin{document}

\title{Theoretical Implications of Diffuse Non-Thermal Emission from
Clusters of Galaxies}

\author{Torsten A. En{\ss}lin}
\affil{Max--Planck--Institut f\"ur Astrophysik,
Karl-Schwarzschild-Str.1, 85740 Garching,  Germany} 

\begin{abstract}
A short review on theoretical implications of non-thermal emission
(radio, extreme ultraviolet, high energy X-ray) from the intra-cluster
medium is given. The origin of cluster radio halos and cluster radio
relics is discussed within the framework of a network of processes
producing a non-thermal electron population. Emphasis is given to the
role of old, remnant, presently invisible relativistic plasma released by
former radio galaxies.
\end{abstract}
\section{Introduction}
The space between galaxies in clusters of galaxies is filled with a
hot and dilute thermal gas, which is the main component of the
intra-cluster medium (ICM).  Radio astronomical observations gave the
first evidence for the occurrence of diffuse non-thermal phenomena in
the ICM with the detection of the radio halo of the Coma cluster
(Willson 1970). Recently more non-thermal signatures were detected
from Coma and a number of other clusters with the observations of
extreme ultra-violet (EUV) emission (Lieu et al. 1996) and high
energy X-ray (HEX) emission (Fusco-Femiano et al. 1999) in excess of
the thermal one. The interpretation of these fluxes is still not yet
settled, nor are the different classes of diffuse radio emission
completely understood. We give an overview on current interpretations
and theories. This should outline the physical insight we will gain
from understanding the currently strongly increasing number of
non-thermal detections of cluster of galaxies (Feretti 1999).

\section{Diffuse Radio Emission from Clusters of Galaxies}

Radio observations are the most sensitive probes of non-thermal
processes in clusters of galaxies, and especially low frequency radio
astronomy plays a crucial role since the radio spectra are steep.  The
diffuse radio sources can be divided into two classes: cluster radio
halos and cluster radio relics\footnote{{\it `Cluster radio relics'}
should not be confused with dying radio galaxies, which are also
called {\it `radio relics'}. For this reason Luigina Feretti proposed
to call them {\it `peripheral halos'}, and Ron Ekers proposed {\it
`radio flotsam'}. Since all three names have disadvantages I decided
to stick to the old one.}.

The morphology of the radio halos is similar to that of the X-ray gas.
They are unpolarized and occur preferentially in clusters which also
show evidence for ongoing or recent cluster merger events.  Cluster
radio relics are irregularly shaped, more peripherally located with
respect to the cluster center and sometimes linearly polarized.  Both,
cluster relics and halos, seem to be connected to cluster merger, but
their natures are different.

\section{Cluster Radio Relics}

There is increasing evidence that the cluster relics appear at
locations where shock waves of an ongoing merger event, or from the
accretion flow onto the cluster are located. In the cases of Abell
2256 and Abell 1367 temperature substructures of the hot ICM gas could
be detected (Briel \& Henry 1994; Donnelly et al. 1998), which support
the presence of a shock wave at the location of cluster relics in
these clusters. For Abell 2256, Abell 3667 and also the Coma cluster
numerical simulations of merger events were satisfactorily fitted to
the X-ray data, which also supports the shock wave-relic connection
(Roettiger, Burns, \& Pinkney 1995; Roettiger, Burns, \& Stone 1999).

The shock wave probably accelerates electrons (and also protons) via
Fermi acceleration. The accelerated electrons illuminate the cluster
magnetic fields by synchrotron radiation leading to the cluster
relic. The magnetic fields get compressed and therefore aligned with
the shock plane, causing the radio emission to be polarized if seen
edge on.  Thus, the observed properties of cluster radio relics, such
as as degree and direction of polarization, surface luminosity,
peripheral position etc., can be understood (En{\ss}lin et al. 1998;
Roettiger, Burns, \& Stone 1999).

But shock waves are common due to the high merging fraction of cluster
of galaxies, whereas cluster relics are only rarely found in
clusters. This could be a surface brightness problem, or the
indication of a second necessary condition (in addition to a shock
wave) for the occurrence of a cluster relic.

The cluster relic 1253+275 in Coma might tell us what this second
condition is: its radio emission is morphological connected to the
radio tails of the galaxy NGC 4789 (Giovannini, Feretti, \&
Stanghellini 1991). Following the radio tails of that galaxy from the
head to the cluster relic, one sees clear evidence for spectral
steepening of the electron population in the radio spectral index, and
a flattening at the location of the relic. This suggests that it is
the radio plasma the galaxy releases which actually flashes up at the
shock wave and not the normal inter-galactic medium (IGM). Radio
plasma has the ideal composition for this: evidently magnetic fields
and a pre-accelerated electron population are present. It is not
necessary for a cluster radio relic, as it is the case for 1253+275,
that the source of the radio plasma is nearby or visible, since also
old invisible radio plasma sitting in the IGM could be re-activated by
shock waves.

\section{Radio Ghosts}
Active radio galaxies fill large volumes in the IGM with radio plasma.
This plasma becomes rapidly invisible to radio telescopes due to
inverse Compton (IC) and synchrotron energy losses of the relativistic
electrons. Afterwards it might form an invisible, but possibly
important phase of the IGM.  Due to the difficult detectability the
name {\it `radio ghosts'} for patches of this remnant radio
plasma\footnote{More nomenclature: `Since the term `ghost' does not
fit the context from the viewpoint of Hindu philosophy, Gopal Krishna
proposed {\it `radio fossil'}, drawing upon the Indian mythological
tale of lady Ahilya (ref: the epic {\it Ramayana}, circa 2000
B.C.). Turned into a fossil while meditating, Ahilya was revived to
life at the touch of Lord Rama's feet, as the Lord passed her way on a
pre-ordained track during his 14-year exile in the forest. Likewise,
`radio fossils' can spring back to a radiant life upon passage of a
shock wave.}  was proposed (En{\ss}lin 1999).

After injection of the radio plasma it will expand or contract until
it reaches pressure equilibrium with the surrounding medium. Its
pressure is given by that of the confined relativistic particles and
the magnetic fields, assumed to be in rough energy equipartition.
Therefore magnetic field energy density in radio ghosts should be
lower, but of the order of the thermal energy density of the
environment.  Subsonic turbulence in this environment, which has an
energy density below the thermal energy density, is therefore not
strong enough to overcome the magnetic elastic forces of the radio
ghost. Sonic or super-sonic turbulence, which is e.g. expected in
giant merger events of cluster of galaxies, can `shred' the ghost into
smaller pieces. The size of such pieces will be comparable to the eddy
size of the turbulence. Since a typical turbulent spectrum has less
energy density on smaller scales there is a length-scale below which
the turbulence is not able to overcome magnetic forces.  Turbulent
erosion of radio ghosts should stop at this length-scale, leaving
small-scale patches of still unmixed old radio plasma.

Radio ghosts are practically invisible as long as their electron
population remains at low energies. But if the population is
re-accelerated the ghost becomes radio luminous again. This can happen
when the ghost is dragged into a large-scale shock wave, e.g. in a
merger event of clusters of galaxies or at the accretion shock where
the matter is falling onto a cluster. The emission region is expected
to be irregularly shaped, and should exhibit linear polarization due
to the compression of the magnetic fields in the shock. Such regions
are indeed observed at peripheral locations of a few clusters of
galaxies: the cluster radio relics.  Further observable signatures of
radio ghosts are discussed in En{\ss}lin (1999), Medina Tanco \&
En{\ss}lin (2000), and En{\ss}lin \& Kaiser (2000).

\begin{figure}
\plotone{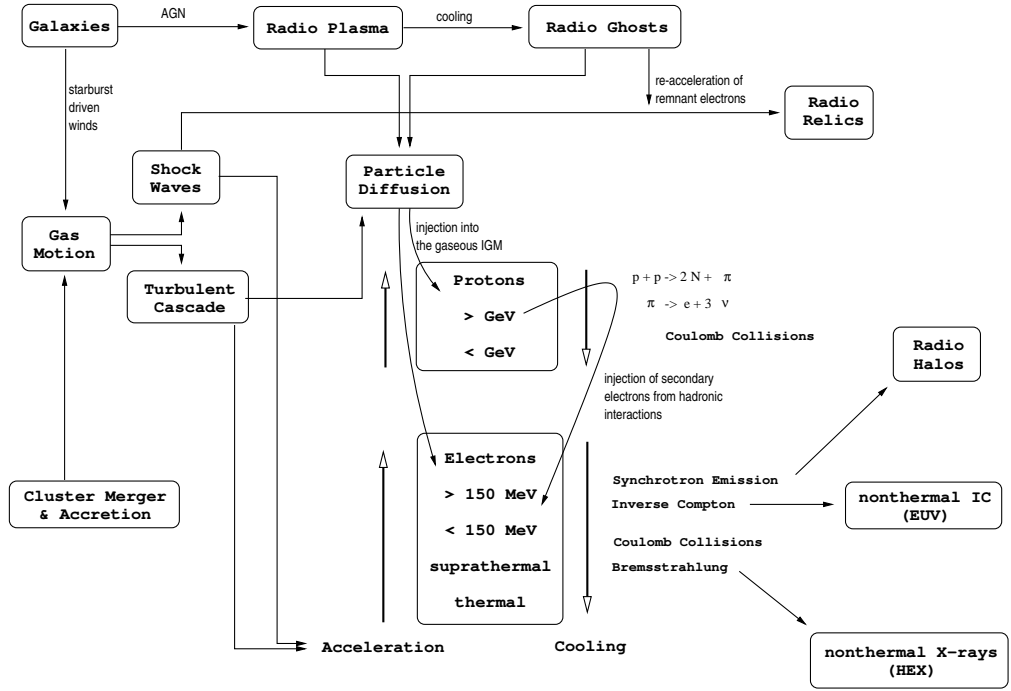}
\caption[]{\label{fig:diag} Network of processes possibly responsible
for non-thermal cluster emission. Description in the text.}
\end{figure}

\section{Cluster Radio Halos\label{sec:halo}}

The origin of the relativistic electrons and magnetic fields in the
ICM, which are necessary for the synchrotron process which produces
the radio halo, is not known. The magnetic fields could be primordial
(Olinto 1997), injected by galactic winds (Kronberg, Lesch, \& Hopp
1999; V\"{o}lk \& Atoyan 1999), or by radio galaxies (Daly \& Loeb
1990; Okoye \& Onuora 1996), produced in shock waves of the large
scale structure formation (Kang, Rachen, \& Biermann 1997), amplified
in the turbulence from galaxy motion (Roland 1981; de Young 1992;
S\'{a}nchez-Salcedo, Brandenburg, \& Shukurov 1998) and cluster
formation (Roettiger, Stone, \& Burns 1999; Dolag, Bartelmann, \&
Lesch 1999), or of another unknown origin. Harris \& Miley (1978) and
Komissarov (1985) proposed that radio halos are superpositions of
large numbers of relic radio galaxis.

There are also many possible sources for the relativistic
electrons. Fig. \ref{fig:diag} gives an overview about the current
theories and their dependencies. In the following the different parts
of the diagram are explained and references to the literature are
given:

The energy source of all non-thermal processes in clusters should be
either the kinetic energy of matter falling onto clusters, or the
outflows from galaxies. The latter can be divided in galactic winds,
which are strongest for starburst galaxies, and ejection of radio
plasma from an AGN. All these processes can produce shock waves and
inject turbulence into the ICM, and therefore produce conditions where
Fermi mechanisms accelerate particles (Jaffe 1977; Schlickeiser, Sievers, \&
Thiemann 1987; Roettiger, Burns, \& Loken 1993; Tribble 1993;
V{\"o}lk, Aharonian, \& Breitschwerdt 1996; Deiss et al. 1997; Eilek
\& Wetherall 1999; Wetherall \& Eilek 1999).

The non-thermal particles could be accelerated out of the thermal pool
(Liang 1999), or they were produced by radio galaxies (Jaffe 1977;
Rephaeli 1977; 1979; Valtaoja 1984). In the latter case, they had to
escape out of the radio plasma, which is difficult since the necessary
diffusion across magnetic fields is a extremely slow
process. Turbulence in the ICM can strongly increase that diffusion
coefficient, and especially shortly after mayor merger events there
exists a short time window when particles might be able to escape
(En{\ss}lin 1999).

The non-thermal electron population produces bremsstrahlung, inverse
Compton and synchrotron emission, which allows us to detect its
presence, but also leads to energy losses which have to be compensated
by re-acceleration (Jaffe 1977; Giovannini et al. 1993; Brunetti et
al. 1999) or injection of fresh electrons. If the radio electrons are
injected and not accelerated in-situ, the sources have to be
distributed throughout the cluster volume. Otherwise, if only a few
located sources produce the electrons, cooling times of the electrons
are much shorter than plausible travel times from the source to the
boundaries of the observed radio halos (Jaffe 1977; but see Holman,
Ionson, \& Scott 1979; Ragot \& Kirk 1997). A promising injection
mechanism is therefore secondary particle production from hadronic
interactions of relativistic protons with the background gas (Dennison
1980; Vestrand 1982; Blasi \& Colafrancesco 1999):
\begin{eqnarray}
  \label{eq:pp}\nonumber
  p + p &\rightarrow& 2 \,N + \pi^\pm \\
  \pi^\pm &\rightarrow& \mu^\pm + \nu_{\mu}/\bar{\nu}_{\mu} \rightarrow
  e^\pm + \nu_{e}/\bar{\nu}_{e} + \nu_{\mu} + \bar{\nu}_{\mu}\nonumber
\end{eqnarray}
The lifetime of relativistic protons in the ICM is of the order of the
Hubble time, or larger (Berezinsky, Blasi, \& Ptuskin 1997). Thus they
are able to travel large distances from their sources before they
release their energy. The production of electrons via charged pions
has to be accompanied by gamma ray production via neutral pions
(Vestrand 1982; En{\ss}lin et al. 1997; Colafrancesco \& Blasi 1998;
Blasi 1999):
\begin{eqnarray}
  \label{eq:ppgamma}
  p + p &\rightarrow& 2\, N + \pi^o \nonumber\\
  \pi^o &\rightarrow& 2 \, \gamma \nonumber
\end{eqnarray}
Thus clusters with radio halos might have gamma-ray halos, which
would be, if detected, a direct proof for a hadronic origin of radio
halos.

\section{The Electron Spectrum\label{sec:e}}

\begin{figure}[t]
\plotone{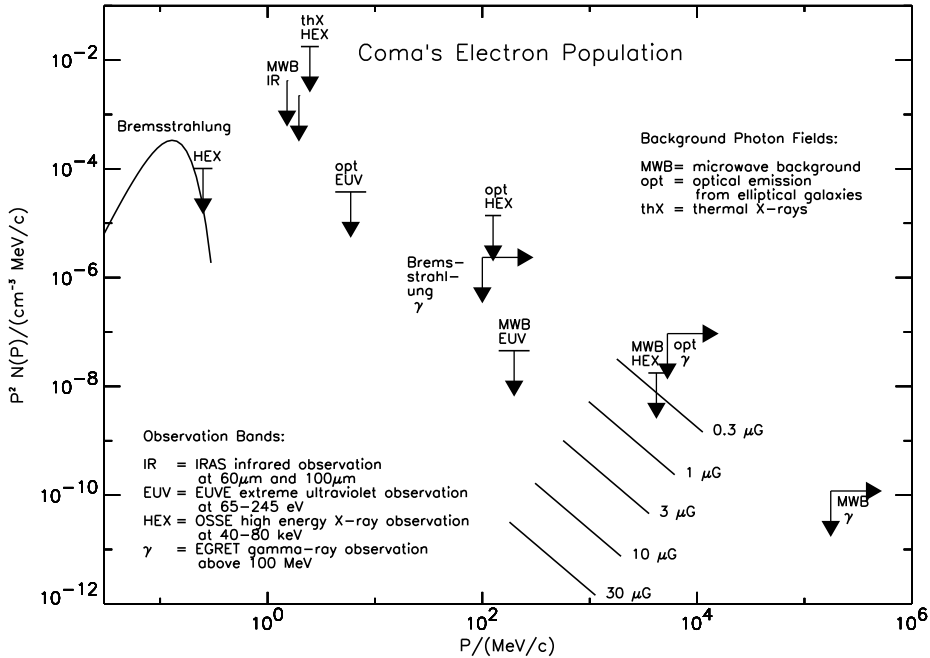}
\caption[]{\label{fig:e} The electron spectrum in the center of the
Coma cluster (from En{\ss}lin \& Biermann 1998). The solid line below
1 MeV$/c$ is the thermal electron spectrum and the lines around 1 GeV$/c$
give the radio emitting electrons for several magnetic field
strength. The upper limits result from several upper limits to
possible IC and bremsstrahlung processes.}
\end{figure}

The electron spectrum is formed by various injection, cooling and
acceleration processes, which should leave their fingerprints on it.
Fig. \ref{fig:e} shows a compilation of the electron spectrum in the
Coma cluster. Note, that the EUV and the HEX upper limits actually
result from detections of excess emissions (compared to the thermal
X-ray emission at these energies; Lieu et al. 1996; Fusco-Femiano et
al. 1999), so that some of the upper limits actually could be data
points.

{\bf The EUV-excess} could result from a low temperature gas (Lieu et
al. 1996). But this is problematic, since one would expect the
presence of resonance lines, which are not detected (Dixon, Hurwitz,
\& Ferguson 1996). Therefore an IC origin of the excess is very
likely. The scattered photon field could be the starlight of the
galaxies, which could explain the narrow radial profile of the
emission, but this would require a very high relativistic electron
energy density in Coma (En{\ss}lin, Lieu, \& Biermann 1999). Probably
the excess is due to IC scattered cosmic microwave background (CMB)
photons (Hwang 1997; En{\ss}lin \& Biermann 1998; Bowyer \&
Bergh\"ofer 1998), which would require a strong electron population
around 150 MeV, not unreasonable since cooling is slowest at this
energy (Sarazin \& Lieu 1998). In order to connect this data point
with the radio electrons a huge step in the electron spectrum is
required, if the Faraday rotation determined magnetic field strength
of $6 \,\mu$G (Feretti et al. 1995) in Coma is correct. An electron
population which was accelerated e.g. in a violent merger event in the
past would have produced such a step due to cooling (En{\ss}lin et
al. 1999; Sarazin 1999; Atoyan \& V{\"o}lk 1999). If the magnetic
fields are inhomogeneous, the required step can be further reduced,
since cooling produces a electron distribution spatially
anti-correlated with the magnetic fields, thus hiding the EUV
producing electrons from radio observations.

{\bf The HEX-excess} can not result from IC scattered starlight
photons, since this would violate other limits to the electron
spectrum (see Fig. \ref{fig:e}). In principle it could result from IC
scattered CMB photons, but this would require magnetic field strength
of $0.16\,\mu$G since otherwise the synchrotron emission of the
responsible electrons would exceed the observed radio halo of Coma
(Fusco-Femiano et al. 1999). This is in strong contradiction to
Faraday rotation measurements. Therefore it is very likely, that a
supra-thermal electron population produces the HEX excess by
bremsstrahlung (En{\ss}lin et al. 1999; Sarazin \& Kempner 1999). Such
a population is expected due to turbulent particle acceleration of
thermal electrons (Dogiel 1999; Liang 1999; Blasi 2000), and it can be
confirmed with future CMB telescopes due to its non-thermal IC
signature (En{\ss}lin \& Kaiser 2000).

\acknowledgments I like to thank the scientific and the local
organizing committee for the invitation to this stimulating
conference. I am grateful for financial travel support from the {\it
International Astronomical Union} (IAU) and the {\it Deutsche
Forschungsgemeinschaft} (DFG). During this journey I enjoyed the
friendly hospitality at the {\it National Center for Radio
Astrophysics (NCRA)} in Pune, at the {\it Inter University Centre for
Astronomy and Astrophysics} (IUCAA) in Pune, and the {\it Raman
Research Institute} (RRI) in Bangalore.

\end{document}